# Giant impact on early Ganymede and its subsequent reorientation

Authors

**Naoyuki Hirata** [a, *]

* Corresponding Author E-mail address: hirata@tiger.kobe-u.ac.jp

**Authors' affiliation**

[a] Graduate School of Science, Kobe University, Kobe, Japan.

Editorial Correspondence to:
Dr. Naoyuki Hirata
Kobe University, Rokkodai 1-1 657-0013
Tel/Fax +81-7-8803-6566

**The origin and early evolution of the Jovian moon Ganymede, known to have an internal ocean[1,2], have garnered considerable interest in the field of origin of satellites and life. Ganymede has an ancient impact structure, called a furrow system[3-7]. The furrow system is the largest impact structures in the outer solar system and the impact should have significantly affected Ganymede's early history; however, its impact is poorly understood. Here we show that mass redistribution induced by the furrow-forming impact caused a reorientation (true polar wander) of Ganymede. The center of the furrow system is located close to the tidal axis, indicating that the impact created a positive mass anomaly that reoriented the impact site toward the tidal axis. We found that an impactor with a radius of 150 km and an incidence angle between 60° and 90°can reproduce the current location of the furrow system. Furthermore, this ejecta model is adoptable in Pluto's reorientation. Although it is proposed that Pluto's reorientation indicates the presence of a global ocean[8,9], our model indicates that it occurs even if no ocean.**

Ganymede is the largest satellite in the solar system and has many unique features, including tectonic troughs known as furrows (so-called the Galileo-Marius furrow system)[3-7]. Furrows are remnants of an ancient giant impact, extending concentrically from a single point of Ganymede, 21° S 179° W (Fig. 1), however estimating the size of the furrow-forming impactor is complicated because of the absence of an identifiable clear rim[10-13]. They are morphologically similar to the multi-ring impact structures on Europa or the Valhalla impact basin on Callisto[3,14,15]. Furrows are the oldest surface features recognized on Ganymede because they are crosscut by impact craters with diameters exceeding 10 km (ref.[5]). Therefore, furrows can provide a window into the early history of Ganymede and have had a largest-scale impact on the outer solar system soon after satellite formation.

We propose that the ejecta mass of the furrow-forming impact created a large positive gravity anomaly around the impact center, followed by the reorientation of Ganymede. Previous studies have not paid attention to the fact that the center of the furrow system coincides with the longitude of Ganymede's tidal axis; however, this coincidence implies that Ganymede has experienced a significant reorientation. Interestingly, the center of the furrow system is very similar in geometric location to the center of the Sputnik Planitia (18° N 178° E [7,16]), the largest impact basin on Pluto. Both centers are located along the tidal longitude and deviated poleward by 20° from the tidal axis. Note that the coordinates of Ganymede and Pluto are defined such that the tidal axes are at latitude of 0° and longitudes of 0° and 180°. Previous studies have proposed that the Sputnik Planitia has a large positive gravity anomaly to cause significant reorientation of Pluto[8,9,16].

In general, a positive gravity anomaly on the surface of a tidally locked satellite leads to the reorientation of a satellite, wherein the anomaly approaches the tidal axis; alternatively, a negative gravity anomaly leads to reorientation, wherein the anomaly approaches the rotation axis[17-20]. Crater holes and ejecta blankets act as negative and positive gravity anomalies, respectively. Both anomalies mostly cancel each other out, but the overall gravity anomaly becomes slightly negative, leading to the net effect of reorientation of the crater center toward the rotation axis[21]. In fact, some of the largest impact basins, such as the south-pole Aitken Basin (Moon) and Rheasilvia Basin (Vesta), are located close to the polar region. In contrast, the Sputnik Planitia basin is located near the tidal axis. Previous studies[8,9] have proposed that isostasy can be achieved within the basin floor if Pluto has a global ocean, which, together with the nitrogen deposits within the basin, crates a large positive gravity anomaly. Because the isostatically compensated topography (e.g., supported by the buoyancy of the asthenosphere or subsurface ocean) has no free-air gravity anomaly, a negative gravity anomaly of the basin depression vanishes even if it has a large positive/negative profile.

A similar mechanism is appropriate for the case of the furrow-forming impact on Ganymede because the formation of multi-ring basins on the Jovian icy satellites has been explained by the collapse of a crater basin owing to a fluid or warm convective ice beneath a thin ice shell[14,22-25]. Although it is difficult to argue whether Ganymede had a differentiated interior with an ocean or un-differentiated interior without ocean at the formation of the furrow system, numerical simulations[26-29] have shown that both a thin ice shell over the ocean and warm convective ice without ocean can reproduce the shallowing of the basin floor and widening of the crater rim of the multi-ring basin. Isostasy would be achieved within the basin floor in both subsurface models (Methods). Then, the ejecta blanket alone played the role of a positive gravity anomaly, leading to the reorientation toward the tidal axis.

We evaluated the extent to which ejecta mass was sufficient to induce significant reorientation in Ganymede. The dimensionless parameter $Q$, defined as the ratio of the degree-2 gravitational potential perturbation of the load to the rotational bulge, is useful in determining solutions[30,31]. If the load is too small compared with the existing bulge (i.e., $|Q| \ll 1$), reorientation does not occur because the bulges play a role in stabilizing the current rotation axis. Here, $Q > 0$ ($Q < 0$) indicates that the load plays the role of a positive (negative) gravity anomaly, and $|Q| = 1$ indicates that the gravity anomaly of the ejecta mass is comparable with that of the rotation bulge. Additionally, we considered the lithospheric thickness because the degree of isostatic compensation depends on it. Various estimations exist regarding the thickness of the lithosphere at the time of furrow formation: (i) the thickness of the lithosphere was estimated as 6 to 10 km based on the width and spacing of furrows[32-34]; and (ii) the effective elastic thickness was estimated as 0.5 km based on the flexural uplift around furrows[35]. Fig. 2 shows $Q$ as a function of the volume of ejecta blankets described by the transient crater radius (Methods). If the central region of the furrow system was not isostatically compensated, we can obtain $Q < 0$. If the lithospheric thickness is $t =$10 km, the volume of ejecta blankets that create a gravity anomaly with $Q = 1$ corresponds to 100% of ejecta created by a transient crater with a radius of 600 km. Conversely, when $t =$0.5 km, resulting in $|Q| \ll 1$, the degree of reorientation is very small. The reorientation of Ganymede does not occur when $t =$0.5 km.

The location of the Sputnik Planitia, which deviates 20° from the tidal axis, was explained by the balance between the bulges and mass anomaly of the basin[8]. However, the center of the furrow system cannot be explained in the same manner, because Ganymede should attain its new orientation roughly 1000 years after the furrow-forming impact while the remnant bulges of Ganymede should be readjusted less than one year (Methods). In this case, the remnant bulges eventually vanish with the occurrence of reorientation; a new bulge is immediately formed in response to the new rotation state, and the gravity anomaly ultimately migrates to the tidal axis[17]. We calculated the center of the gravity anomaly (i.e. the minimum principal axis of the load) of mass distribution of the ejecta blanket created by the furrow-forming impact. Fig. 3 and Extended Data Fig. 1 show examples of the global distribution of an ejecta blanket before (left plates) and after (right plates) the most stable reorientation. The distribution of an ejecta blanket before reorientation is computed by the ejecta trajectory method[36] including incidence angle (Methods). Consequently, we found that an impactor suitably reproduces the location of the furrow system regardless of the initial location. Using the Monte Carlo method, we examined the probability that the center of the furrow system after the most stable reorientation matches the current location (Methods). For example, the probability that the impact center falls within 5° from the point deviated poleward by 20° from the tidal axis under the conditions of impact incidence angle between 60° and 90°, transient crater radius of 700 km, and the ejecta launch angle of 30° is 57.2% (Extended Data Fig. 4). If reorientation does not occur, the probability is 0.76%. The probability tends to be the maximum when the size of transient crater radius is 700 or 800 km, which agrees with the above arguments for $Q$ value. The transient crater radius of 700 or 800 km is produced by an impactor with a radius of 150km (ref.[13]). However, it should be noted that the center of the furrow system can be explained by not only the ejecta trajectory method including incidence angle but also inhomogeneous compensation on the surface; for example, inhomogeneous lithospheric thicknesses leads to an asymmetric load, even if the ejecta blanket is perfectly symmetric. Then, the ejecta blanket rim supported by the thicker lithosphere would be the center of the gravity anomaly, similar to examples in Fig. 3. Nonetheless, conclusively determining this hypothesis is difficult because the thickness distribution four billion years ago was highly uncertain, although the lithospheric thicknesses of the Galileo and Marius Regios were estimated 10 and 6 km, respectively[32-34].

The location of the Sputnik Planitia can be explained in the same manner. The Sputnik Planitia appears nitrogen deposits within its basin floor. The surface of the nitrogen deposits is ~2 km deep relative to the mean radius[37], the density of the nitrogen deposits and water ice are 1000 and 930 kg/m$^3$, respectively, and the thickness of the nitrogen deposits is 10 km (ref.[38]). Those observations indicate that the nitrogen deposits offset 90% of the magnitude of the mass anomaly of the topographic depression of the basin. Therefore, it is possible that the ejecta blanket is a primary sources of a gravity anomaly. If the center of the load of the Sputnik Planitia is determined by the ejecta blanket alone, the probability defined above is 42.5% when the conditions of impact incidence angle between 60° and 90°, transient crater radius of 300 km, and the ejecta launch angle of 30° (Methods and Extended Data Fig. 5). Because our ejecta model does not require the subsurface ocean, Pluto's reorientation occurs even if Pluto does not have an ocean. It is known that a convective ice shell without an ocean is also plausible in Pluto, according to some of thermal models predicted by ref.[39]. Although Pluto's reorientation was the only evidence for the existence of an ocean, our model indicates it is likely that Pluto has no ocean.

The reorientation of Ganymede provides a window into the early history of the Jovian satellites. The amount of ejecta in the Valhalla basin on Callisto is equivalent to $Q = 5.6$, which is sufficiently large to cause a reorientation of Callisto (Methods); however, the Valhalla is far from both the pole and tidal axes. Possible explanation includes that Callisto's rotation at the formation of the Valhalla basin has been faster than that at present. If Callisto's rotation period was less than 50 hours, the ejecta mass from the Valhalla basin would not have led to the reorientation of Callisto. This is not unlikely because it is estimated that Callisto attains a synchronously rotating state (a rotation period of 400 hours) in 220 million years after its formation[40] and the Valhalla is one of the oldest surface features on Callisto[23,41]. It is proposed that the 10-km-thick lithosphere of Ganymede indicates a surface heat flow of approximately 40 mW/m$^2$ at the time of furrow formation[34], which exceeds the maximum plausible heat fluxes expected from radiogenic heating alone at 4.5 billion years ago (nearly 27 mW/m$^2$)[42]. Therefore, it may indicate additional energy sources, such as tidal heating, gravity segregation, and/or loss of accreational heat[35] at the time of the furrow formation. Perhaps, the flexural uplift around the furrows, indicating a thin lithosphere ($t = 0.5$ km), may reflect the elastic thickness during a geologically active period of Ganymede long after furrow formation, as pointed out by ref.[35]. A large amplitude of libration or non-synchronous rotation state of Ganymede should have lasted 1000 years after the furrow-forming impact. In contrast, the mass anomaly of the ejecta blankets supported by the lithosphere would vanish for less than one billion years, as indicated by the evidence of viscous relaxation of old craters[42]. The cooling of Ganymede during the timescale has possibly increased the thickness of the lithosphere and should have fixed the orientation of Ganymede. It is known that heliocentric impactors preferentially hit the leading hemisphere of a tidally locked satellite, and the impact crater density of Ganymede decreases from the apex (center of the leading hemisphere) to the antapex (center of the trailing hemisphere) of the motion; however, the degree of the apex-antapex asymmetry in various tidally locked satellites is considerably lower than that of theoretical estimates for ecliptic comets[43-45]. Previous studies[43-45] have proposed various possible explanations for this difference, such as crater saturation, non-synchronous rotation, reorientation, and nearly isotropic comets as the dominant impactor population. At least, further reorientation would be unlikely in the case of Ganymede, although reorientation by flipping the current north and south poles or shifting the longitude by 180° is possible.

Many areas of Ganymede still have not been imaged with sufficient resolution[7], and further data from future explorations are required for further discussions of the tectonic landforms formed by reorientation and the age of the furrow formation. For example, tectonic patterns (other than furrows) resulting from Ganymede's reorientation have not been discovered, although tectonic patterns resulting from the reorientation of Pluto and the Jovian satellites have been investigated[9,46,47]. It is possible that new fractures may not have formed if re-movement of existing fractures, or furrows, absorbs reorientation-induced stresses. The current accuracy of the gravity and topography measurements of Ganymede[48-50] is insufficient for the comparison between individual topographic features owing to the lack of a global shape model of Ganymede and a local topographic map around the center of the furrow system. The gravity anomaly may have vanished, but the topographic profiles of the ejecta blankets should still be visible in the topography even if they are mostly viscously relaxed. Future explorations would reveal such a remnant of topographic profiles associated with the furrow-forming impact and the reorientation of Ganymede, which would provide insights into Ganymede's early history and highlight its differences compared with other Jovian satellites.

**Main Reference**

**Figure Legends**

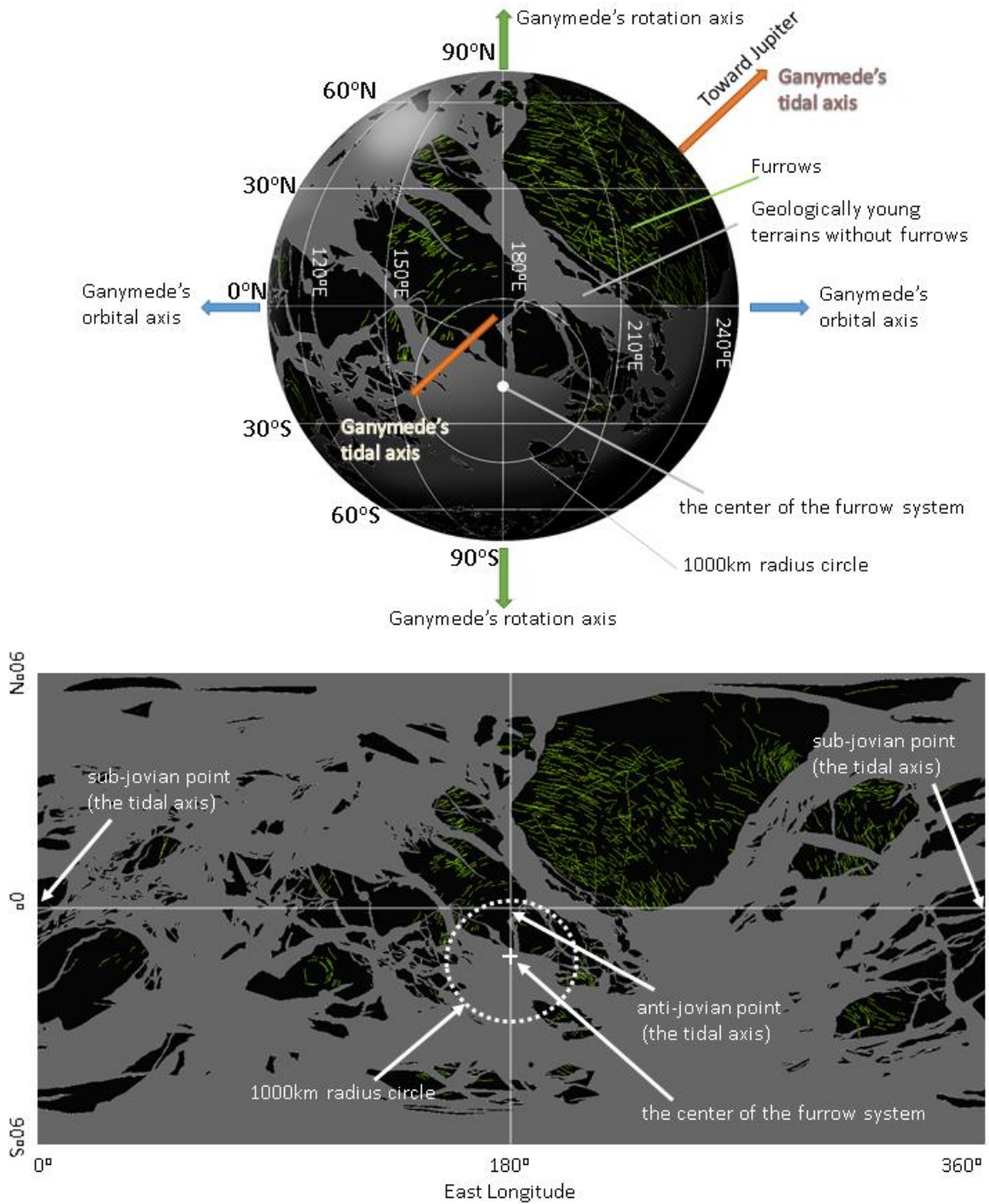

**Figure 1.** The distribution of furrows and location of the center of the furrow system shown in the hemisphere that always faces away from Jupiter (top) and the cylindrical projection map of Ganymede (bottom). The distribution of furrows was obtained from ref.[12]. Gray regions show geologically young terrain without furrows. Furrows (green lines) exist only on geologically old terrains (black regions).

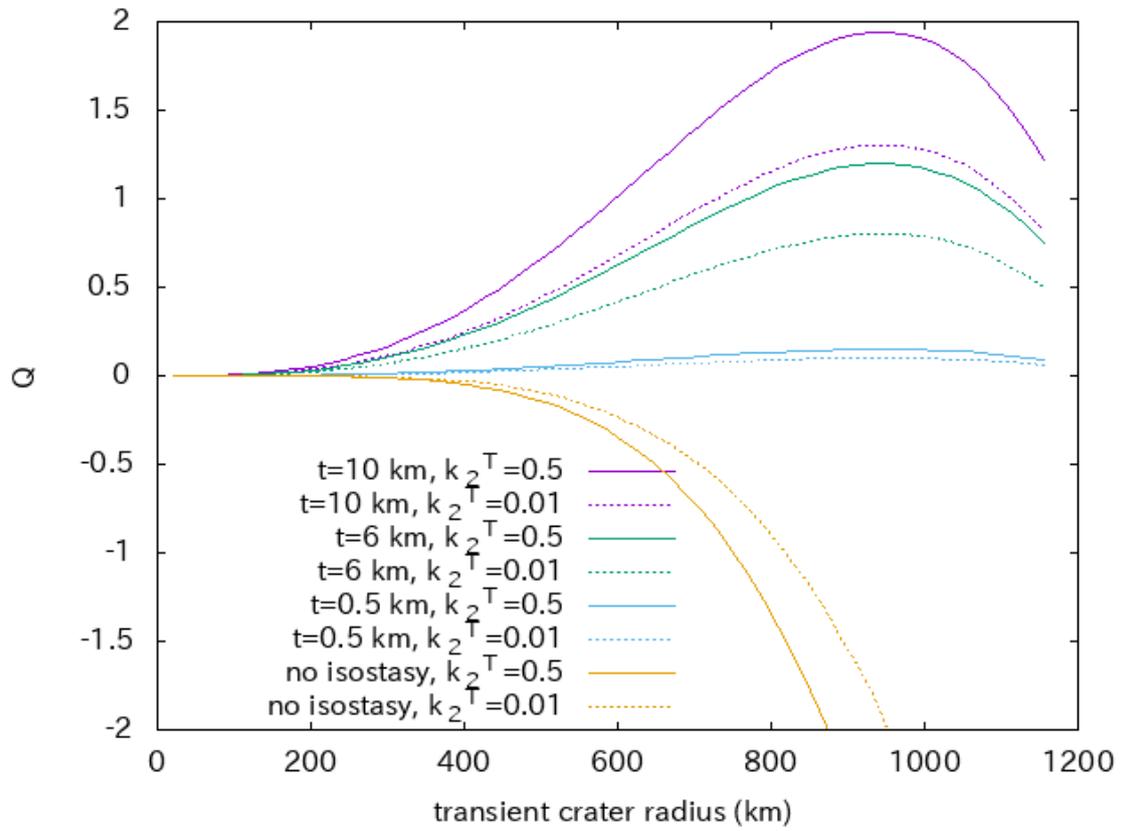

**Figure 2.** The parameter of $Q$ as a function of the transient crater radius, where $t$ and $k_2^T$ represent the lithospheric thickness and tidal Love number, respectively.

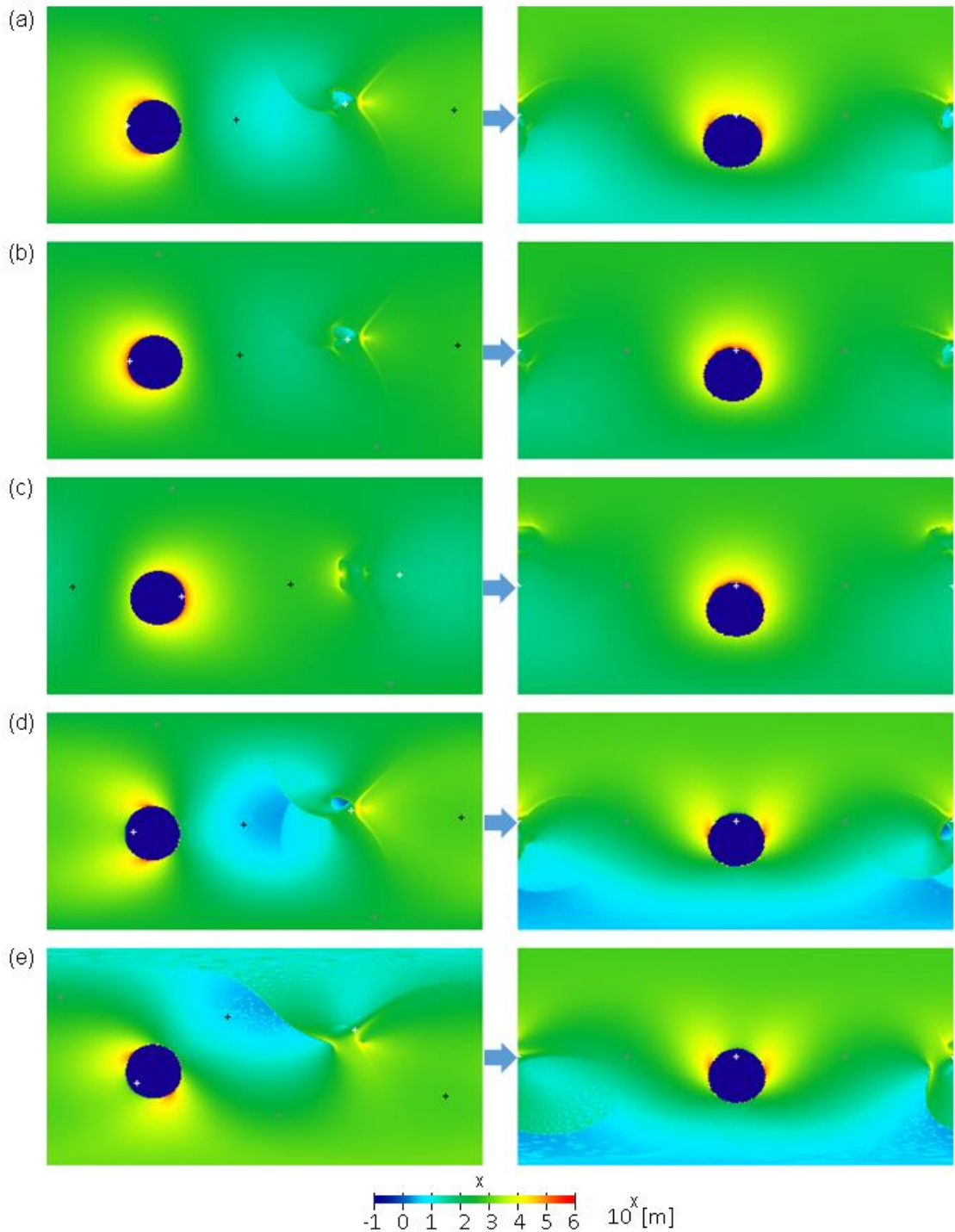

**Figure 3.** (left) Global distribution of ejecta blankets before reorientation using impact parameters of #1-5 presented in Extended Data Table 1. White, gray, and black "+" symbols represent the minimum, intermediate, and maximum principal axes, respectively. (right) The global ejecta thickness after reorientation to ensure that Ganymede had the most stable rotation. Vertical and horizontal axes in each plate indicate the latitude and east longitude, respectively, where the top left corner is 90° N and 0° E, and the bottom

right corner is 90° S and 360° E. The color bar is represented in a logarithmic scale.

**Methods**

**The dimensionless parameter, $Q$**

Many significant studies have been conducted on the reorientation of solid bodies of the solar system[18,19,20,46,47,51,52]. For a rigid-spinning tidally-locked body, mass redistribution, such as cratering or subsurface diapirism, leads to a new principal axis rotational state (reorientation), while the remnant rotational and tidal bulges stabilize the rotation axis. Therefore, the true polar-wander solutions are determined by the balance between the remnant bulges and an uncompensated load subsequent to mass redistribution[30,31]. The dimensionless parameter $Q$ for a rigid-spinning body is expressed as follows[30,31]

$$Q \equiv \frac{3GM\sqrt{5}g_{20}}{\Omega^2 R^3 (k_2^{T*} - k_2^T)}, \tag{1}$$

where $g_{20}$ is the degree-2 order-0 coefficients of the gravitational potential perturbation of the mass redistribution at the object surface; $G$ is the gravitational constant; $M$, $R$, and $\Omega$ are the mass, radius, and rotation angular frequency of Ganymede, respectively; and $k_2^{T*}$ and $k_2^T$ are the degree-2 fluid and actual Love numbers, respectively. Although the $k_2^T$ of Ganymede at 4 billion years ago is highly uncertain, it is approximately 0.01 if present-day Ganymede has no ocean and its mantle has a high viscosity of $\eta > 10^{14}$ Pa s, or approximately 0.4 to 0.6 if Ganymede has an ocean or its mantle has a low viscosity of $\eta < 10^{12}$ Pa s (ref.[53]). We studied the two cases of $k_2^T = 0.01$ and $k_2^T = 0.5$.

Following ref.[21], we assumed the following conditions for a crater basin: (i) the ejecta blanket of excavated material has a uniform thickness, $t_e$, and a simple annulus shape with inner and outer radii of $R\phi$ and $R\theta$, respectively, where $\phi$ and $\theta$ are the angles from the center of the object, respectively, and the outer radius is twice the inner radius, $\theta = 2\phi$; and (ii) the basin floor has an uniform depth, $h$, and a simple circular shape with a radius of $R\phi$. In this case, a simple geometric solution can be used to determine the volume of the ejecta blanket and hole as follows:

$$V_{ejecta} = 2\pi R^2 t_e (\cos\phi - \cos\theta), \text{ and} \tag{2}$$

$$V_{hole} = 2\pi R^2 h (1 - \cos\phi). \tag{3}$$

In general, both volumes are equal to each other because of the conservation of mass[21], even if the very shallow basin was created after the collapse of a transient crater; subsequently, we can use the simple relation of $h = t_e(\cos\phi - \cos\theta)/(1 - \cos\phi)$, if isostasy is not achieved within the basin floor. We considered the variations in the density of a nearly spherical object expressed by variations over a series of concentric spherical shells, and the density perturbations were azimuthally symmetric. The density perturbation in this ejecta blanket load model is expressed as:

$$\rho(r) - \rho_0(r) = \begin{cases} \begin{cases} 0 & (\phi > \varphi > 0) \\ \rho_c & (\theta > \varphi > \phi) \\ 0 & (2\pi > \varphi > \theta) \end{cases} & (R < r < R + t_e) \\ \begin{cases} -\rho_c & (\phi > \varphi > 0) \\ 0 & (2\pi > \varphi > \phi) \end{cases} & (R - h < r < R) \\ 0 & (r < R - h) \end{cases}, \quad (4)$$

where $\rho_0(r)$ is the mean density for radius $r$, $\rho_c$ is the density of the crust, and $\varphi$ is the colatitude from the crater center. As the degree-2 order-0 coefficient of the density perturbation is expressed by[54]

$$\rho_{20}(r) = \frac{\sqrt{5}}{4} \int_0^\pi \rho(r,\varphi)(3\cos^2\varphi \sin\varphi - \sin\varphi) \, d\varphi, \quad (5)$$

we can obtain

$$\rho_{20}(r) = \begin{cases} \frac{\sqrt{5}}{4}\rho_c(\cos\theta \sin^2\theta - \cos\phi \sin^2\phi) & (R < r < R + t_e) \\ -\frac{\sqrt{5}}{4}\rho_c \cos\phi \sin^2\phi & (R - h < r < R) \\ 0 & (r < R - h) \end{cases}. \quad (6)$$

The degree-$l$ order-$m$ spherical harmonic coefficients of the gravitational potential at an arbitrary distance, $s$ ($s \geq R$), is given by[54]:

$$g_{lm}(s) = \frac{4\pi}{M(2l+1)s^l} \int_0^R r^{l+2} \rho_{lm}(r) \, dr, \quad (7)$$

where $s$ is a distance from the center.

Therefore, if this density perturbation is supported by an infinitely rigid lithosphere, the degree-2 order-0 gravitational potential coefficient at the object surface was described by

$$g_{20} = \frac{\pi R^2 \rho_c}{\sqrt{5}M}[t_e \cos\theta \sin^2\theta - (h + t_e)\cos\phi \sin^2\phi]. \quad (8)$$

When isostasy is not achieved within the basin floor (i.e. $h = t_e(\cos\phi - \cos\theta)/(1 - \cos\phi)$), the gravitational coefficient is always negative: $g_{20} < 0$.

    A mass excess/deficit that is compensated for by isostasy does not create a gravity anomaly, while the ejecta mass supported by the bending and membrane stresses of the lithosphere creates a gravity anomaly. Therefore, the gravity anomaly reduces by a factor of $(1 - C_n)$, where the degree of isostatic compensation, $C_n$, was described by[55]

$$C_n = \left[1 - \frac{3\rho_m}{(2n+1)\bar{\rho}}\right]\left[\frac{\sigma(N^3 - 4N^2) + \tau(N-2) + N - (1-\nu)}{N - (1-\nu)} - \frac{3\rho_m}{(2n+1)\bar{\rho}}\right]^{-1}, \quad (9)$$

where $N = n(n + 1)$, $n$ is the degree of spherical harmonics of the horizontal width of the load ($n = 2\pi R/w$, where $w$ is the horizontal width of the load), $\bar{\rho}$ is the mean density of the satellite, and $\tau$ and $\sigma$ are nondimensional parameters defined as follows:

$$\tau \equiv \frac{Et}{R^2 g(\rho_m - \rho_c)} \quad \text{and} \quad \sigma \equiv \frac{\tau}{12(1-v^2)}\left(\frac{t}{R}\right)^2, \tag{10}$$

where $t$ is the elastic thickness, $\rho_m$ is the density of the mantle underlying the lithospheric plate, $E$ and $v$ are Young's modulus and Poisson's ratio of the lithosphere, and $g$ is the surface gravity. Here, $C_n = 0$ indicates that the load is fully supported by the lithosphere and $C_n = 1$ indicates that the load is fully compensated by isostasy. Although the degree of isostatic compensation depends on the strength of the lithosphere, we assume $E = 9$ GPa and $v = 0.32$ (ref.[56]). We used $\rho_m = 1000$ kg m$^{-3}$ and $\rho_c = 930$ kg (ref.[57]) as the differences in density. As the ejecta blanket spreads around a couple of basin radii, we assumed $w = 1000$ km. Note that the degree of isostatic compensation does not significantly differ even if we use $w = 500$ to $3000$ km. Consequently, we obtained $C_n = 0.87$ for $t = 10$ km, $C_n = 0.92$ for $t = 6$ km, and $C_n = 0.99$ for $t = 0.5$ km.

The numerical simulation[28] or impact experiment[58] proposed that the impactor can penetrate Europa's ice shell and create conduits to the underlying ocean. Similarly, if Ganymede has an ocean and its ice shell is sufficiently thin, isostasy is easily achieved within the basin floor by the inflow of water. The furrow-forming impact alone may not create a large ocean, because the size of a local melt pool created by the impact is up to 6 times impactor radii[59], which is less than 10 % of a total volume of Ganymede. However, isostasy would be achieved within the basin floor even if Ganymede's interior is warm convective ice without ocean. Numerical simulations for warm convective ice without ocean[13,29] shows that the cold lithospheric material is replaced by warm subsurface material and the post-impact lithospheric thickness becomes quite thin within the transient crater radius, when the transient crater size is sufficiently larger than the lithospheric thickness. The lithospheric thickness to recover by cooling from the surface is roughly determined by the characteristic thermal diffusion depth, $\sqrt{\kappa t_d}$, where $\kappa$ is the thermal diffusivity and $t_d$ is a timescale[60]. The relaxation time of a load with a horizontal width of 100 km is 25 years (see next section). If we use $\kappa = 2.9 \times 10^{-5}$ for ice[61,62] and $t_d = 25$ years, we can obtain $\sqrt{\kappa t_d} = 150$ m. A elastic thickness of $t = 150$ m does not support a load above the lithosphere at all ($C_n = 1$). The mass anomaly within the basin floor vanishes quickly before a recovery of the lithospheric thickness. Therefore, isostasy can be achieved within the basin floor of the furrow system, regardless of the two surface models. If only the basin floor is isostatically compensated for ($C_n = 1$), the gravity anomaly in the basin floor is removed, which has the same attribute as $h = 0$. In this case, the gravity anomaly becomes positive ($g_{20} > 0$).

Thus, the dimensionless parameter becomes

$$Q = \frac{3G\rho_c(1-C_n)}{\Omega^2 R(k_2^{T*}-k_2^T)} \frac{V_{ejecta}}{2R^2}(\cos^2\phi + \cos\phi\cos\theta - \sin^2\theta). \quad (11)$$

Based on the Z model[63], we assumed that the total volume ejected from a crater ($V_{ejecta}$) was a fraction of the total volume displaced from the transient crater ($V_{tc}$), $V_{ejecta} = \frac{1}{4}V_{tc}$, and that the size of the transient crater was a simple hemisphere, $V_{tc} = \frac{2}{3}\pi r_{tc}^3$. Additionally, we defined the radius of the isostatically compensated region to be equivalent to the apparent crater rim radius, $R\phi = 1.3 r_{tc}$. As a result, we obtained the value of $Q$ as a function of the transient crater radius, as shown in Fig. 2. In Fig. 2, we assume $\theta = 2\phi$ based on ref.[21]. For the no isostasy case in Fig. 2, we assumed $h = t_e(\cos\phi - \cos\theta)/(1-\cos\phi)$ and $t = 10$ km ($C_n = 0.87$).

The radius of the transient crater of the Gilgamesh basin is 135 km (ref.[64]), which indicate a load of $Q\sim0.001$. The Gilgamesh basin hardly reorient Ganymede, although a reorientation caused by the Gilgamesh basin has previously been proposed[46]. As another example, the mound at the sub-Jovian point, discovered by the Juno flyby and legacy data reanalysis, with a height of 3 km and an oval of 450 km × 750 km (refs.[65,66]), has a value of $Q=0.05$ ($h=-3$ km, $t_e= 0$, and $R\phi=300$ km), even when assuming the most optimistic case, $C_n = 0$ and $k_2^T =0.5$. Following Eq. (41) in ref.[18], the mound leads to very small reorientation of 0.3° if its latitude is 45°.

**Tidal decay of rotation**

Following a large impact event, tidal friction quickly dampens the motions of the satellite's orientation, such as the librational oscillations and non-synchronous rotation, and the satellite attains its new orientation on a short timescale[21]. The timescales of the tidal decay of librational and non-synchronous rotation have been estimated in previous studies[21,40,67], and all three literatures were equivalent. Following ref.[67], the timescale is expressed as

$$T_{damp} = \frac{2}{3}\frac{Q_d}{k_2^T}\frac{GC}{n_o^3 R^5}, \quad (12)$$

where $C$ is the maximum moment of inertia about the center of mass; $n_o$ is the mean orbital motion; and $Q_d$ is the specific dissipation factor of the satellite. Assuming a homogenous interior ($C = 0.4\ MR^2$) and reasonable values for warm icy satellites, $k_2^T=0.5$ and $Q_d=100$ (ref.[68]), we can obtain $T_{damp}=870$ years. If Ganymede was cold and rigid (say $Q_d/k_2 \sim 10^5$), we can obtain $T_{damp}=4.4$ million years.

The mass anomaly not supported by the lithosphere disappears owing to viscous relaxation on a timescale of $\eta/\rho_m g w$ (ref.[60]), where $w$ is the horizontal width of the load. Icy materials underlying the elastic layer are convecting, and their reference viscosity is as low as $\eta = 10^{12}$ to $10^{17}$ Pa s (ref.[69]). For example, a 100 km-width load not supported by the lithosphere vanishes in less than 25 years. Similarly, the density perturbations below the lithosphere created by heating and redistribution of the impact cratering would vanish on a short timescale. If the relaxation time of the remnant tidal/rotational bulges is given by $\eta/\rho_m g R$ (ref.[19]), the bulges would vanish in 1 year. Because these timescales are much shorter than those of the tidal decay of rotation, both (i) the mass anomaly not supported by the lithosphere and (ii) the tidal/rotational bulges would not contribute to Ganymede's final orientation. The gravity anomaly of the ejecta blankets supported by the lithosphere should eventually vanish, but its timescale would be sufficiently longer period than 1000 years. Theoretically determining its timescale is difficult because it depends on heat fluxes and a thermal conductivity. Nonetheless, because it is estimated that old craters on the dark terrain of Ganymede is relaxed viscously in 1 billion years[42,70], the timescale of relaxation of the mass anomaly would be also approximately or less than 1 billion years.

**The Valhalla basin**

It is known that the crater density in the outer graben zone of the Valhalla basin is intermediate between that of the inner zones and unmodified cratered terrain and apparently increasing linearly outward, which would be due to obliteration by continuous ejecta blanket[5,23]. The original crater radius of the Valhalla basin was estimated to be 500 km based on the mapping of ejecta and secondaries[71]. The thickness of the lithosphere was estimated to range from 15 to 20 km during the formation of the Valhalla basin[14]. Assuming $R\phi=500$ km, $k_2^T = 0.5$, $t = 20$ km, $h = 0$, $w = 500$ km, $n = 15$, $C_n = 0.66$, and physical parameters for Callisto in Eq. (12), we can obtain $Q = 5.6$ for the Valhalla basin, which should lead to a significant reorientation of Callisto. Possible explanations for this include Callisto's rotation period and other surface loads. If Callisto's rotation period was 50 h (currently 400 h), we could obtain $Q = 0.09$. This is not unlikely because Callisto attains a synchronously rotating state in 22 million years for a specific dissipation factor of $Q_d = 100$, and 220 million years for $Q_d = 1000$ after its formation[40]. Therefore, Callisto at the formation of the Valhalla basin may not have reached a synchronously rotating state yet. Note that, because Ganymede attains a synchronously rotating state in 0.66 million years for $Q_d = 100$ after its formation[40], Ganymede's rotation period at the time of furrow formation should not much differ from now. Finally, as many areas of Callisto have not yet been imaged with sufficient resolution, we cannot rule out the possibility that large undiscovered loads existed on Callisto's surface.

**Calculation of the ejecta blanket**

Numerical simulations and experimental studies have shown that, although the shape of the crater depression is not sensitive to the impact incidence angle unless the angle is very shallow, the shape of the ejecta blanket is strongly sensitive to the impact incidence angle, and most impacts produce asymmetric ejecta patterns[72-76]. Unless an ejecta blanket is perfectly symmetric, the center of the load does not match that of the crater. To compute the thickness distribution of the ejecta blanket, the methods described by refs.[36,77] were followed. In this study, the mass and initial launch velocity of ejecta particles obtained from refs.[78,79] were used. The ejecta model of ref.[78] was based on an experiment with a normal impact (incidence angle of 90°). Ref.[79] updated the ejecta model by including the incidence angle of the impactor as a parameter. The trajectories of the ejecta particles were solved using Hill's equation[80]. We defined ejecta particles that reached below the object surface as colliding with the object, and ejecta particles that reached an altitude greater than the Hill radius of the object as escaping from the object. Additionally, we removed the particles landing within the apparent crater radius to assume an isostatically compensated basin floor. We calculated the moment of inertia tensor of the mass redistribution of each ejecta blanket, and the eigenvalues and eigenvectors of the tensor to obtain the principal moments of inertia and principal axes of each ejecta blanket. We remapped each ejecta blanket in the body-fixed coordinates after reorientation to ensure that Ganymede had the most stable rotation; the maximum and minimum principal axes of the mass distribution matched the rotation and tidal axis, respectively.

Fifteen examples of the cases with an ejecta launch angle of 45° and C4 (dry sand in the gravity regime) as the set of scaling constants are presented in Extended Data Table 1 and shown in Fig. 3 and Extended Data Figs. 1 and 2. In many cases, we found that the current location of the center of the furrow system could be reproduced (Fig. 3 and Extended Data Fig. 1). However, this could not be reproduced in some cases (Extended Data Fig. 2). Although reproducing the current location of the furrow system involves many parameters and is complicated, our results suggest that (i) the impact incidence angle is between 60° and 90°; or (ii) the uprange direction is roughly on the west side. This is almost independent of the initial location of the impact site (Extended Data Fig. 1). The thickness of the ejecta blankets at the rim is 100 km. However, any subsequent movement after landing is not considered, and in reality, some dispersion induced by mass wasting or re-ejection would occur.

In this ejecta model, a small mound with a height of approximately 10 km owing to the accumulation of ejecta appears near the sub-Jovian point. Therefore, in some cases, the mound appears at the sub-Jovian point after the reorientation, which may agree with the small mound with a height of 3 km located near the sub-Jovian point (1.5° E, 0.5° N) discovered by the Juno flyby and legacy data reanalysis[65,66]. When the uprange direction is roughly on the west side, the mound tends to appear near the sub-Jovian point after the reorientation, because the small mound shifted slightly from the exact opposite point of the impact owing to the effect of Ganymede's rotation. Viscous relaxation, isostatic readjustment, and/or uncertainty of the ejecta model may be responsible for the difference in height and shape between the discovered mound and theoretically evaluated ejecta mound on the opposite side.

Using the Monte Carlo method, we examined the probability that the center of the furrow system after the most stable reorientation moves within 5° or 10° from the point deviated poleward by 20° from the tidal axis. The initial location, uprange direction, and incidence angle of the impactor were generated by uniformly distributed random numbers; by using four uniformly distributed random numbers, $0 \leq x_1, x_2, x_3, x_4 < 1$, the longitude, latitude, uprange direction, and incidence angle follows functions of $2\pi x_1, \mathrm{acos}(2x_2 - 1), 2\pi x_3,$ and, $\mathrm{acos}(\sqrt{x_4})$, respectively[43]. We examined cases for an impact incidence angle between 60° and 90° or between 30° and 60°, a target material of C4 or C1 (water), a transient crater radius of 500, 600, 700, 800, or, 900 km, and an ejecta launch angle of 25°, 30°, or 45° (angle between the initial launch velocity vector and the object surface). Note that ejecta launch angle of ~28° may be the most appropriate in a giant impact because of the curvature of the target surface[81,82]. For each case, 1000 trials were performed. Extended Data Fig. 3 shows the initial and final locations of the crater centers in the case of an impact incidence angle between 60° and 90°, a target material of C4, and the ejecta launch angle of 30°, or 45°. Extended Data Fig. 4 shows the probability as a function of the transient crater radius.

Numerical simulation for Pluto is almost the same as that for Ganymede, other than the trajectories of the ejecta particles, which were solved using the equation of motion for the circular restricted three-body (Pluto, Charon, and an ejecta particle)[80]. Extended Data Fig. 5 shows the same probability as a function of the transient crater radius. Note that the mean radius of the Sputnik Planitia is 550 km (ref.[9]), which corresponds to the transient crater radius of 420 km.

**Code Availability**

The software for calculating ejecta blankets is available at GitHub

(https://github.com/naoyukihirata/ganymede-ejecta).

**Method Reference**

**Acknowledgements**

We appreciate Shunichi Kamata for helpful comments. This study was partly supported by the JSPS Grants-in-Aid for Scientific Research (Nos. 20K14538 and 20H04614) and Hyogo Science and Technology Association.

**Author Contributions**

N.H. carried out all.

**Competing Interest declaration**

None

**Extended Data Table 1.** Parameters for each ejecta blanket

| # | Fig. | Lon.[*1] | Lat.[*1] | uprange direction [*2] | incidence angle [*3] |
|---|------|------|------|-----|----|
| 1 | 3a | 90 | -10 | 0 | 70 |
| 2 | 3b | 90 | -10 | 0 | 80 |
| 3 | 3c | 90 | -10 | 180 | 80 |
| 4 | 3d | 90 | -10 | 0 | 60 |
| 5 | 3e | 90 | -10 | 45 | 60 |
| 6 | EX1a | 180 | -10 | 0 | 70 |
| 7 | EX1b | 280 | -10 | 0 | 70 |
| 8 | EX1c | 180 | -80 | 0 | 70 |
| 9 | EX1d | 270 | -60 | 0 | 70 |
| 10 | EX1e | 90 | -10 | 270 | 80 |
| 11 | EX2a | 90 | -10 | 0 | 90 |
| 12 | EX2b | 90 | -10 | 180 | 70 |
| 13 | EX2c | 90 | -60 | 0 | 70 |
| 14 | EX2d | 90 | -10 | 0 | 50 |
| 15 | EX2e | 90 | -10 | 315 | 30 |

*1 East longitude and latitude of the impact site before reorientation.

*2 Uprange direction in degrees, anticlockwise from the east.

*3 Incidence angle of the impactor from the surface.

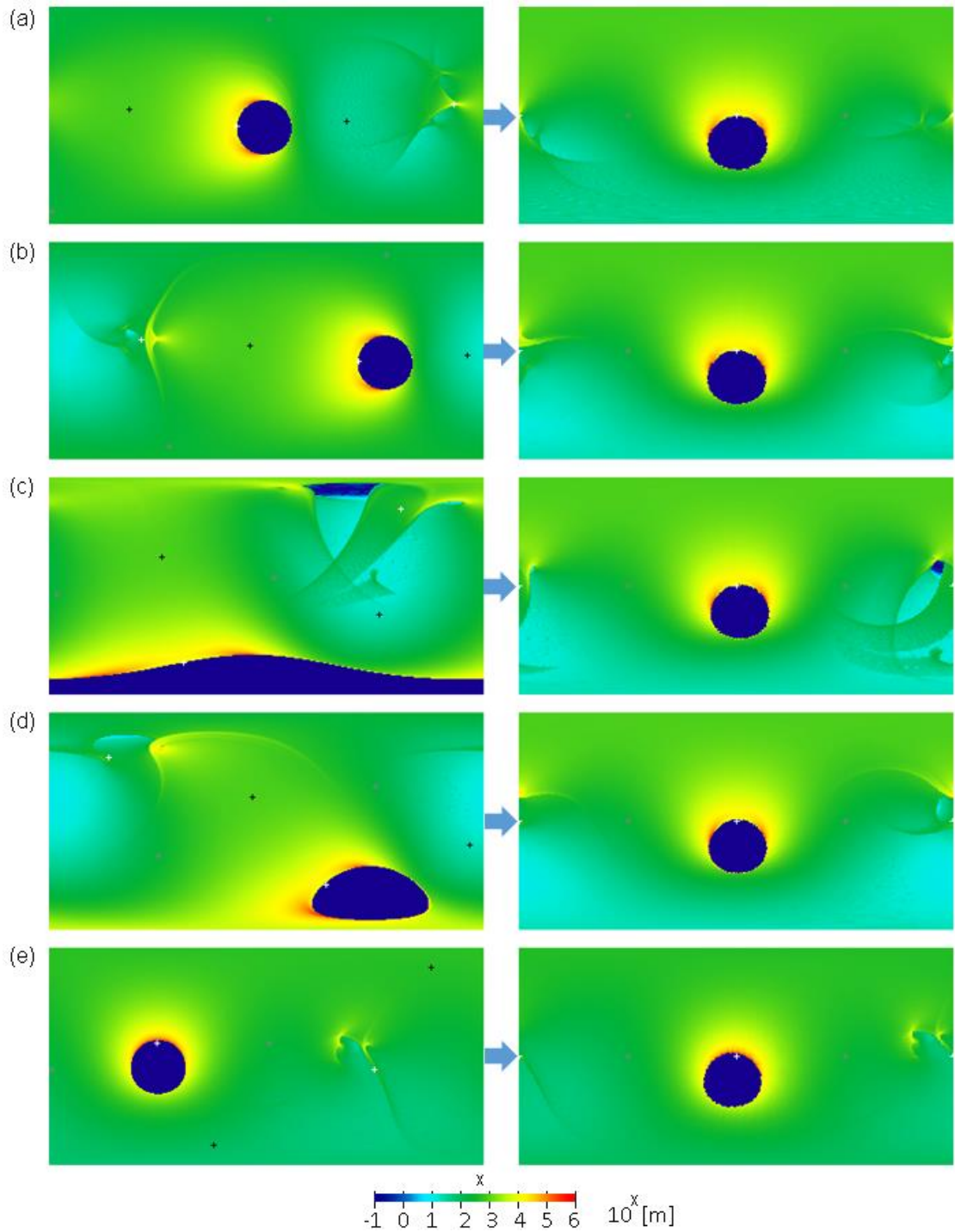

**Extended Data Figure 1.** Global distribution of ejecta blankets before reorientation (left) and after reorientation (right) using impact parameters of #6-10 presented in Extended Data Table 1. White, gray, and black "+" symbols represent the minimum, intermediate, and maximum principal axes, respectively. Vertical and horizontal axes in each plate indicate the latitude and east longitude, respectively, where the top left corner is 90° N

and 0° E, and the bottom right corner is 90° S and 360° E. The color bar is represented in a logarithmic scale.

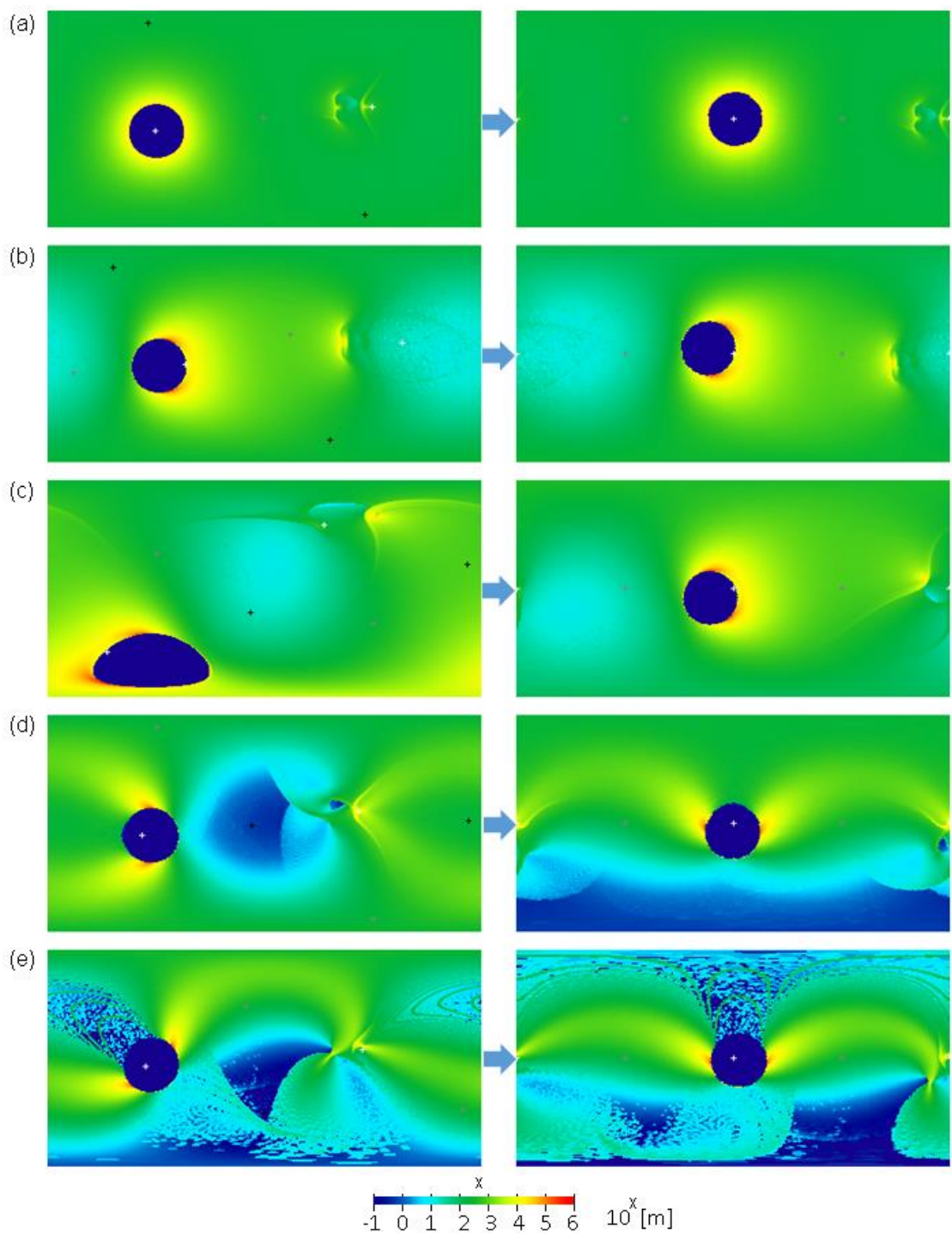

**Extended Data Figure 2.** Global distribution of ejecta blankets before reorientation (left) and after reorientation (right) using impact parameters of #11-15 presented in Extended Data Table 1.

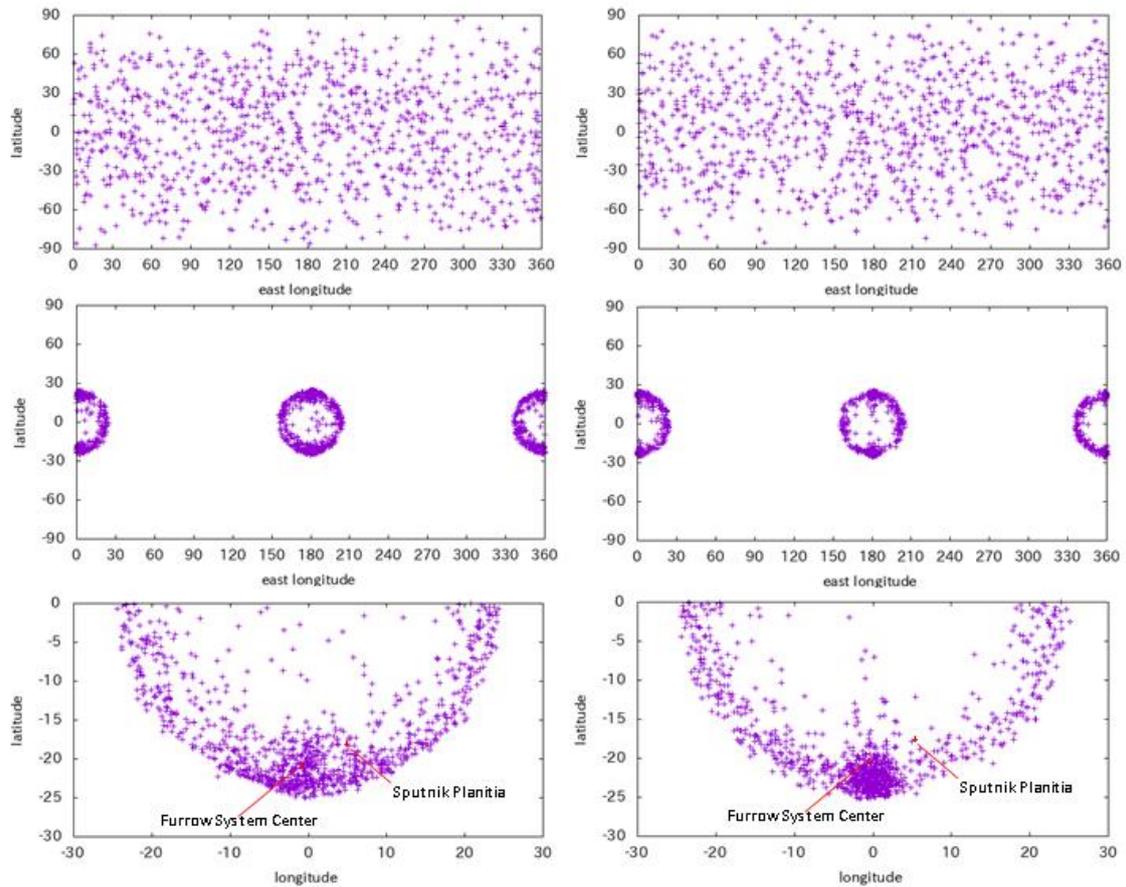

**Extended Data Figure 3.** (top) The initial location of the center of the craters, (middle) the final location of the center of the craters after the reorientation, and (bottom) the final location in the coordinate centered at the tidal axis. We have displayed cases of ejecta launch angles of 45° (left) and 30° (right). Note that because the true polar wander solution does not distinguish between the southern and northern hemispheres, points plotted in the northern hemisphere are shown with the reversed sign. Owing to satellite rotation, it is not symmetrical with respect to the east-west direction.

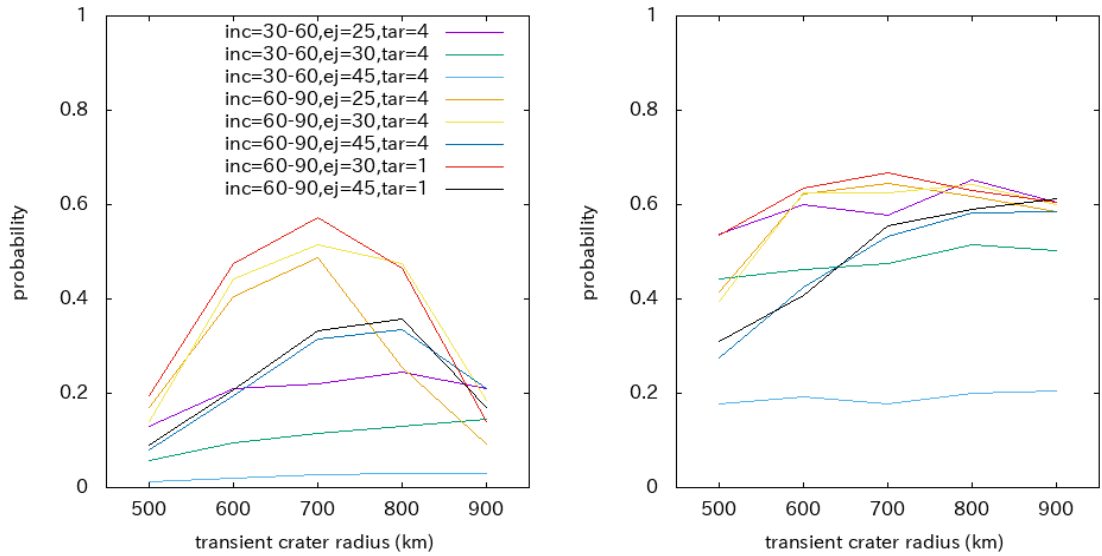

**Extended Data Figure 4.** The probability that the center of the furrow system after the most stable reorientation moves within 5° (left) or 10° (right) from the point deviated poleward by 20° from the tidal axis, as a function of the transient crater radius. Here we assume the impact incidence angle (inc) between 30° and 60° or between 60° and 90°, ejecta launch angle (ej) of 25°, 30°, or 45°, and target material (tar) of C1 or C4. Legend is common on both plates.

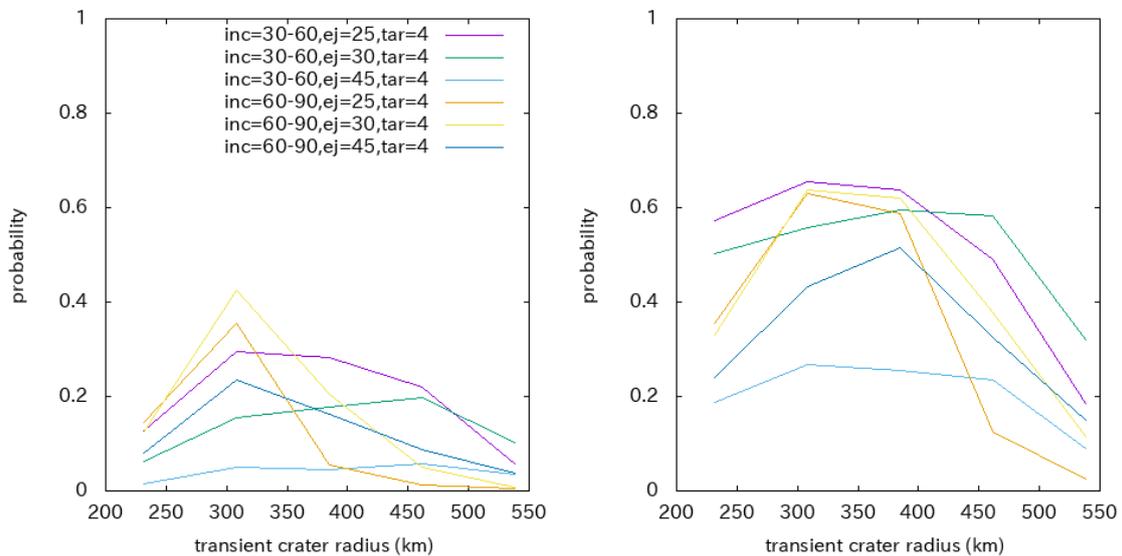

**Extended Data Figure 5.** The probability that the center of the Sputnik Planitia after the most stable reorientation moves within (left) 5° or (right) 10° from the point deviated poleward by 20° from the tidal axis, as a function of the transient crater radius, assuming the impact incidence angle (inc) between 30° and 60° or between 60° and 90°, ejecta launch angle (ej) of 25°, 30°, or 45°, and target material (tar) of C4.